\documentclass[preprint,12pt]{elsarticle}

\usepackage{graphicx}
\DeclareGraphicsExtensions{.png}
\usepackage{amssymb}
\usepackage{amsmath}

\usepackage[nodots]{numcompress}

\usepackage{gensymb}
\usepackage[numbered]{bookmark}
\usepackage{comment}

\usepackage{caption}
\usepackage{subcaption}
\usepackage{float}

\usepackage{setspace}

\biboptions{sort&compress}

\journal{Physica B: Physics of Condensed Matter}

\begin{document}

\begin{frontmatter}



\title{Micromagnetic structures: flexomagnetoelectric coupling point symmetry and Neumann's principle}


\author[a]{B. M. Tanygin}\ead{b.m.tanygin@gmail.com}

\address[a]{Faculty of Radiophysics, Taras Shevchenko National University of Kyiv, 4G, Acad. Glushkov Ave., Kyiv, Ukraine, 03187}

\begin{abstract}
\doublespacing

Recent series of articles (B.M.~Tanygin, 2011-2012) has been aimed to study flexomagnetoelectric properties of magnetically ordered crystals by means of a magnetic point symmetry description. Besides its fundamental interest (complete symmetry classification), applications to a classical micromagnetic variational problem of a thermodynamically equilibrium state calculation had been provided without a proof. It was shown that the magnetic point symmetry group determination of the given micromagnetic structure could be used to predict \textit{qualitatively} the electric polarization distribution induced by a flexomagnetoelectric coupling. This article is an important addendum which provides a formal mathematical explanation why and how a symmetry-based description of flexomagnetoelectric phenomena conforms to a \textit{quantitative} calculation of an equilibrium state of a magnetization and electrical polarization vectors spatial distribution. An application of the Neumann's principle and a group-theoretical methodology to real micromagnetic structures is clarified as well. The well-known role of a symmetry-based approach in new interactions and phenomena discoveries has been discussed.

\end{abstract}


\end{frontmatter}



\section{Introduction}
\label{ref_intro}
\doublespacing

A group-theoretical approach was originally well applied to ferroelectric domain structures~\cite{Janovec1976}. A magnetic point symmetry classification of different micromagnetic structures had been established~\cite{Bar'yakhtar1984,Bar'yakhtar1986,Bar'yakhtar1984a,Privratska1999,Privratska2002,PrIvratska2007,Tanygin2009,Tanygin2011a,Tanygin2011,Tanygin2012a,Tanygin2012,Tanygin2012d}: the complete list of magnetic point groups is obtained for magnetic domain walls, Bloch lines, Bloch points, magnetic vortex and skyrmion structures. An important practical feature of this symmetry classification is its ability to predict \textit{qualitatively}~\cite{Bar'yakhtar1986,Privratska1999,Tanygin2011a,Tanygin2011} a magnetization $\textbf{M}$ and/or electric polarization $\textbf{P}$ vector equilibrium spatial distribution: an even/odd/zero/other type of components $M_i\left(r_j\right)$ and $P_i\left(r_j\right)$ had been identified for each point group.

Hence, a knowledge of the magnetic point group $G_{\mathrm{P}}$ of a crystal sample simultaneous paramagnetic and paraelectric phase~\cite{tanygin2010magnetic} allows to describe (bypassing calculations) equilibrium spatial distributions of magnetization and polarization with~\cite{Tanygin2011a} or without~\cite{Tanygin2009} their magnetoelectric coupling. Although this statement is very natural in the framework of the group-theoretical consideration, it requires a proof in the scope of the micromagnetics~\cite{brown1963micromagnetics} and the Ginzburg~-~Landau theory~\cite{chandra2007landau}. Also, quantitative parameters cannot be obtained from a symmetry theory. The \textit{quantitative} calculation of an equilibrium state of a magnetization and electrical polarization vectors spatial distribution is a solution of micromagnetic problem, which is based on a free energy functional minimization~\cite{brown1963micromagnetics,chandra2007landau}. On the other hand, the inhomogeneous magnetoelectric coupling~\cite{Bar'yakhtar1983} (also known as flexomagnetoelectric (FME) coupling~\cite{Pyatakov2009}) free energy density expression is defined by a crystal symmetry~\cite{Tanygin2012b}. Hence, two approaches are obviously related: a variational problem solution and a symmetry-based prediction. This relation was shown only for some micromagnetic structures without a general proof.

The derivation of this proof as well as a methodological analysis is the purpose of this work. As a constructive example, only Bloch lines in a planar domain wall has been considered. A generalization to other micromagnetic structures is straightforward.

\section{Theory}
\label{ref_symth}

A calculus of variations of a total free energy functional $\Phi$ is given by:
\begin{align} \label{eq:var_problem}
	\delta \int [
	              &F_{\mathrm{M}}\left( \textbf{r},\textbf{M},\nabla_{\mathrm{x}} \textbf{M}, \nabla_{\mathrm{y}} \textbf{M}, \nabla_{\mathrm{z}} \textbf{M} \right) + \nonumber\\
	            + &F_{\mathrm{P}}\left( \textbf{r},\textbf{P},\nabla_{\mathrm{x}} \textbf{P}, \nabla_{\mathrm{y}} \textbf{P}, \nabla_{\mathrm{z}} \textbf{P}  \right) + \nonumber\\
	            + &F_{\mathrm{ME}}\left( \textbf{r},\textbf{M},\textbf{P},\nabla_{\mathrm{x}} \textbf{M}, \nabla_{\mathrm{y}} \textbf{M}, \nabla_{\mathrm{z}} \textbf{M} \right) ] \mathrm{d}^3 \textbf{r} = 0,
\end{align}
where $\mathrm{XYZ}$ is the reference Cartesian crystallophysical coordinate system \cite{Zheludev1983}; the ferromagnetic $F_{\mathrm{M}}$, ferroelectric $F_{\mathrm{P}}$, and flexomagnetoelectric $F_{\mathrm{ME}}$ \cite{Tanygin2012b} components of the local free energy density is integrated over the whole crystal sample volume. The equation~(\ref{eq:var_problem}) could be supplemented by additional conditions, like the equality $\left|\textbf{M}\right| = M_\mathrm{S}$ for a ferromagnetic material. This problem conventionally reduces to a set of Euler-Lagrange equations which can be solved numerically in the general case~\cite{Fischbacher2007}. The FME interaction free energy density (here and hereinafter, Einstein notation is being used):
\begin{equation} \label{eq:fme_energy}
	F_{\mathrm{ME}} = \gamma_{ijkl} P_i M_j \nabla_k M_l
\end{equation}
is an invariant of a magnetic point group of a spatially unlimited crystal in a paramagnetic and paraelectric phase $G_{\mathrm{P}}^{\infty}$ \cite{Bar'yakhtar1983,Tanygin2012b}:
\begin{equation} \label{eq:fme_energy_sym}
	g \cdot \gamma_{ijkl} P_i M_j \nabla_k M_l = \gamma_{ijkl} P_i M_j \nabla_k M_l,
\end{equation}
Here and hereinafter, a magnetic point symmetry transformation $g \in G_{\mathrm{P}}^{\infty}$ is applying to individual vectors and an operator $\boldsymbol{\nabla}$, which is treated as a polar time-even vector. The FME interaction tensor $\gamma_{ijkl}$ is a material constant. A total free energy minimization in scope of the variational problem~(\ref{eq:var_problem}) leads to the following expression~\cite{Mostovoy2006,Betouras2007}:
\begin{equation} \label{eq:polarization}
	P_i\left(\textbf{r}\right) = - \chi_\mathrm{e}\left(\textbf{r}\right) \gamma_{ijkl} M_j\left(\textbf{r}\right) \nabla_k M_l\left(\textbf{r}\right),
\end{equation} where $\chi_{\mathrm{e}}\left(\textbf{r}\right)$ is a dielectric susceptibility of a crystal paramagnetic phase.

In case of a spatially restricted crystal sample, the magnetic point group $G_{\mathrm{P}}$ of a crystal sample in the simultaneous paramagnetic and paraelectric phase~\cite{tanygin2010magnetic} is derived by means of the Curie symmetry principle:
\begin{equation} \label{eq:sample_Curie}
	G_{\mathrm{P}} = G_{\mathrm{P}}^{\mathrm{\infty}} \cap G_{\mathrm{S}},
\end{equation}
where $G_{\mathrm{S}}$ is a magnetic point group of a crystal sample shape. In case of a thin film or bulk plate, it is given by the limit Curie group supplemented by a time-reversal operation $1'$:
\begin{equation} \label{eq:G_S}
	G_{\mathrm{S}} = \infty/\mathrm{m m m} 1'
\end{equation}

The symmetry transformations $\tilde{g} \in U_l$ of a Bloch line magnetic point group $U_l$:
\begin{equation} \label{eq:mm_group}
	U_l \subset G_{\mathrm{P}}
\end{equation}
correspond to functional dependencies of order parameters \cite{Bar'yakhtar1984a,Bar'yakhtar1986,Tanygin2012d}:
\begin{align}
	\tilde{g} \cdot \textbf{M}\left(\textbf{r}\right) = \textbf{M}\left(\tilde{g} \cdot \textbf{r}\right) \label{eq:sym_class_def_M} \\
	\tilde{g} \cdot \textbf{P}\left(\textbf{r}\right) = \textbf{P}\left(\tilde{g} \cdot \textbf{r}\right) \label{eq:sym_class_def_P}
\end{align}
The detailed algorithm and the table data of the magnetic point group derivation depending on the boundary conditions and the domain wall plane as well as Bloch line orientation had been published before~\cite{Tanygin2009,tanygin2010magnetic,Tanygin2012a}. According to (\ref{eq:sample_Curie}) and (\ref{eq:mm_group}~-~\ref{eq:sym_class_def_P}), the crystal sample free surfaces symmetry ($G_{\mathrm{S}}$) perturbs the spatial distribution of electric polarization and magnetization vectors. Although a magnetic point symmetry of this perturbation can be the same for the same experimental sample shape like (\ref{eq:G_S}), an ab initio mechanism of this impact can be different: the strain-induced magnetocrystalline anisotropy, the demagnetization and depolarization phenomenon, the surface anisotropy~\cite{Aharoni1987}, the boundary conditions of inhomogeneous magnetostrictive and piezoelectric effects, the non-local magnetoelectric coupling~\cite{Bar'yakhtar1984a}, etc.

As a rule, each dependency $M_i\left(r_j\right)$ and $P_i\left(r_j\right)$ (any combination: $i~=~x,y,z$ and $j~=~x,y,z$) is a sum $\left(F\right)$ of an even $\left(S\right)$ and an odd $\left(A\right)$ function (the notation ~\cite{Bar'yakhtar1984,Tanygin2012a} is being used) in the triclinic-pedial ($G_{\mathrm{P}}^{\mathrm{\infty}} = 1'$) magnetically ordered crystal. In case of an arbitrary crystal, aside of the transformation $1'$, additional crystallographic magnetic point symmetry transformations $\tilde{g}$ apply sequentially restrictions (\ref{eq:sym_class_def_M},~\ref{eq:sym_class_def_P}) to the components of the dependencies $\textbf{M}\left(\textbf{r}\right)$ and $\textbf{P}\left(\textbf{r}\right)$ compare to the lowest symmetry case. Possible transformations $\tilde{g}$ are $n$, $n'$, $\bar{n}$, and $\bar{n}'$; where $n \in \left\{1,2,3,4,6\right\}$. This leads to the dependencies $M_i\left(r_j\right)$ and $P_i\left(r_j\right)$ symmetry increase:
\begin{align} \label{eq:sym_increase}
	& \left(F\right) \rightarrow \left(S\right) \nonumber\\
	& \left(F\right) \rightarrow \left(A\right)
\end{align}
Let us consider (here and hereinafter) only classical planar micromagnetic structures (scope of~\cite{Tanygin2012a}), i.e. a planar domain wall with a Bloch line inside: $U_l \subset mmm1'$ \cite{Tanygin2012d}. Then, the group $U_l$ consists of the transformations $\tilde{g}$ with $n =$ 1 or 2 only. Moduli of the vector components $M_i$ and $P_i$ are not changing by such symmetry transformations. For each multi-index $\left(ijkl\right)$ (each index can be $x,y$, or $z$ in the crystallophysical coordinate system \cite{Zheludev1983}), both sides of the equation~(\ref{eq:polarization}) changes sign by the transformation $\tilde{g}$ only simultaneously due to the requirement~(\ref{eq:fme_energy_sym}) taking into account~(\ref{eq:mm_group}).

Hence, the equation~(\ref{eq:polarization}) satisfies the components functional dependence parity requirements (\ref{eq:sym_class_def_M},~\ref{eq:sym_class_def_P}). Former one is a predictive capability in scope of a magnetic symmetry consideration of the given crystal. Latter one is a solution of the FME coupling variational problem~(\ref{eq:var_problem}): a magnetization is treated as a known function and a polarization is treated as an unknown one, which means that a stable micromagnetic structure induces a polarization by FME coupling. Important to note, that any requirements satisfaction by the equation~(\ref{eq:polarization}) does not mean that a solution of (\ref{eq:var_problem}) cannot have other symmetries, which will be considered in the next section.

\section{The Neumann's principle application}
\label{ref_disc}

If some component of $\textbf{M}\left(\textbf{r}\right)$ is zero in terms of a total free energy minimization and its components type of a spatial parity cannot be found in the tables with the magnetic point groups list~\cite{Tanygin2012a} then the Bloch line group $U_l$ should be selected in a way taking into account the rule that a zero functional dependence is a particular case of an even or odd dependency: $(0) \rightarrow (A)$ or $(0) \rightarrow (S)$ or $(0) \rightarrow (F)$ for any component $M_i\left(r_j\right)$ and $P_i\left(r_j\right)$. In this case, the group $U_l$ is a medium symmetry after a spontaneous symmetry breaking $G_P \rightarrow U_l$ through a magnetic ordering; and a magnetic point group $V$ of the variational problem solution $\left[ \textbf{M}\left(\textbf{r}\right),\textbf{P}\left(\textbf{r}\right) \right]$ represents a physical properties symmetry (\textit{cf}. Neumann's principle~\cite{Bhagavantam1967,Post1978}):
\begin{equation} \label{eq:neumann_gr_rel}
	V \supseteq U_l \subset G_{\mathrm{P}}
\end{equation}
Here, the group $V$ describes completely a magnetic point symmetry of equilibrium magnetization and polarization spatial distributions in a framework of a given model (\ref{eq:var_problem}) independently on a crystal sample symmetry $G_{\mathrm{P}}$. As an example of this concept: a magnetic vortices and skyrmion point symmetry has been described by a rotational infinity-fold symmetry axis $\infty_{\mathrm{z}}$ \cite{Tanygin2012} even though this symmetry element does not exist in crystallographic groups \cite{Zheludev1983}.
 
Same methodology should be followed in the case when a particular set of even and odd functions $M_i\left(r_j\right)$ and $P_i\left(r_j\right)$ parity types cannot be found in the tables published in the Ref.~\cite{Tanygin2012a,Tanygin2012d}. According to Neumann's principle, following symmetry reduction is possible in case of group $U_l$ selection: $(A) \rightarrow (F)$ or $(S) \rightarrow (F)$ for any component $M_i\left(r_j\right)$ or $P_i\left(r_j\right)$ in the given Bloch line. It means that an even and odd dependency is a particular case of an arbitrary dependency $(F)$.


It is important to note, that as it was shown in the Ref.~\cite{Bar'yakhtar1986}, if a variational problem solution provides a too highly symmetrical solution then it means that a some free energy invariant has been omitted. Hence, described here methodology of the magnetic symmetry-based predictions application to the real variational problem solutions can be used as a pointer towards the new material properties search.

Let us consider a similar situation in case of a FME coupling. The well-known expression of the FME coupling \cite{Sparavigna1994,Mostovoy2006,Pyatakov2011} is given by a Lifshitz invariant-like single-constant coupling term:
\begin{equation} \label{eq:fme_energy_Lif}
	F_{\mathrm{ME}} = \bar{\gamma}_{ijkl} P_i \left(   M_j \nabla_k M_l - M_l \nabla_k M_j   \right),
\end{equation}
which can be derived from the equation (\ref{eq:fme_energy}) by a removal of total spatial derivatives $\nabla_k M_l M_j$. It was shown \cite{Tanygin2011} that this step cannot be justified by the medium symmetry invariance requirements (\ref{eq:fme_energy_sym}). Hence, a reason of a wide usage of the expression (\ref{eq:fme_energy_Lif}) is related to an experimental experience; i.e. a predominance of a specific microscopical mechanism in case of some published experimental observations of the FME phenomena. A microscopic nature of an exchange-relativistic \cite{Bar'yakhtar1984a} FME interaction has been suggested based on the spin supercurrent model with the similar expression for the cases (different constants) of the superexchange and double-exchange interactions \cite{Katsura2005}: 
\begin{equation} \label{eq:fme_micro_P}
	\textbf{P} = - \Gamma \left[ \textbf{e}_{12} \times \left[ \textbf{S}_1 \times \textbf{S}_2 \right] \right],
\end{equation}
where $\Gamma$ is a material constant; $\textbf{S}_1$ and $\textbf{S}_2$ are spins of transition metal ions in the $M-O-M$ three-atom triad. The $\textbf{e}_{12}$ is a polar time-even unit vector directed along the line segment connected metal ions.

The described type of the FME coupling (\ref{eq:fme_energy_Lif}) requires two non-zero magnetization components. According to the term (\ref{eq:fme_energy_Lif}), magnetic domain walls or spin-density-wave with a collinear magnetic ordering cannot induce an electric polarization. Oppositely, according to the \cite{Tanygin2012a}, if at least one magnetization components $M_\alpha\left(\textbf{r}\right) \neq \textbf{0}$ then nonzero polarization components can be induced by the FME coupling by means of the free energy density term~\cite{Tanygin2012b}:
\begin{equation} \label{eq:M_satur_factor}
f_1 = \gamma_{i \alpha i \alpha} P_i \nabla_i M_\alpha^2 / 2
\end{equation}
in case of $G_{\mathrm{P}}^{\infty}$ belonging to a cubic, orthorhombic system, or tetragonal system (dihedral groups only)~\cite{Tanygin2012b}. A formation of multiferroic phases at the magnetic commensurability transitions of $\mathrm{Y Mn_2 O_5}$ has been explained by the provided FME coupling type (\ref{eq:M_satur_factor}) as a phase dislocation related mechanism \cite{Betouras2007}, which is suppressed by the well-known expression (\ref{eq:fme_energy_Lif}) as a total spatial derivative term \ref{eq:M_satur_factor}. Hence, a progress of an experimental and theoretical research of a novel FME coupling mechanism leads to the same conclusion as a symmetry based one (\ref{eq:polarization}) published originally \cite{Bar'yakhtar1983,Bar'yakhtar1984a}.


Generally, according to the Neumann's principle, a zero value of a specific component $P_\beta\left(\textbf{r}\right)$ is possible in the variational problem solution with given model assumptions. Example is a saturation magnetization $\textbf{M}\left(\textbf{r}\right) = \mathrm{const} \equiv M_{\mathrm{S}}$ constancy in a case of a ferromagnetic material leading to the suppression of the invariant~(\ref{eq:M_satur_factor}) in crystals of a cubic system \cite{Tanygin2012b}.	Also, it is possible in a case of a given measurement tolerance and a macroscopic level of consideration corresponding to a physically infinitesimal volume selection in a research program. Here, if a symmetry-based prediction allows $P_\beta\left(\textbf{r}\right) \neq \textbf{0}$ then a physical mechanism of the FME coupling should be searched: an increase of a variational problem solution accuracy; a considering of all non-zero (allowed by $G_{\mathrm{P}}^{\infty}$) values of material tensor components $\gamma_{ijkl}$; looking for a new integro-differential~\cite{Tanygin2012b,Bar'yakhtar1984a} expressions of a total free energy additionally to the~(\ref{eq:fme_energy_sym}).

The Bloch line always contains direction of a continuous translational symmetry. If $\tilde{g} \cdot \left(e_y \textbf{r}\right) = - \left(e_y \textbf{r}\right)$ and $\tilde{g} \cdot M_i = - M_i$ or $\tilde{g} \cdot P_i = - P_i$ then corresponding components should be zero~\cite{Tanygin2012a}. This is not the case for Bloch points~\cite{Tanygin2012d}, which are more general micromagnetic structures compare to Bloch lines. In other words, a Bloch line is a particular case of a Bloch point. The Neumann's principle application takes place here as well. Magnetic point group $U_l$ of the Bloch point never requires $M_\alpha\left(\textbf{r}\right) = \textbf{0}$ or $P_\beta\left(\textbf{r}\right) = \textbf{0}$. It was shown that that the set of all possible groups $U_l$ (magnetic point symmetry classification) of Bloch points is identical to the one of the Bloch lines~\cite{Tanygin2012d}. There are 48 magnetic point groups of the Bloch points including 22 (11 time-invariant and 11 time-noninvariant ones) enantiomorphic and 26 non-enantiomorphic groups.

\section{Conclusion}
\label{ref_concl}

Hence, symmetry-based predictions of the FME phenomena correspond to a variational problem solution. Neumann's principle corresponds to removal of negligible interaction mechanisms or a model related coarse-grained approximation. It can be treated as a designation for further discovery of new physical interactions in magnetically and electrically ordered media including a coupling of these two subsystems.






\bibliographystyle{model1-num-names}
\bibliography{My-Collection}







\end{document}